\documentclass[journal,onecolumn,12pt]{IEEEtran}

\usepackage{graphicx}
\usepackage{amssymb}
\usepackage{epstopdf}
\usepackage{amsmath}
\usepackage{amsthm}
\usepackage{enumerate}
\usepackage{cite}
\usepackage{mathcomp}
\usepackage{supertabular}
\usepackage{longtable}
\usepackage{stmaryrd}
\usepackage{url}
\usepackage{color}
\usepackage{rotating}
\usepackage{float}
\usepackage[boxed]{algorithm2e}
\usepackage{array}
\usepackage{placeins}
\usepackage{multicol}
\usepackage{setspace}
\usepackage{circuitikz}

\newcommand\remove[1]{}
\newcolumntype{C}[1]{>{\centering\let\newline\\\arraybackslash\hspace{0pt}}m{#1}}

\theoremstyle{definition}

\newtheorem{prop}{Proposition}[section]
\newtheorem{thm}{Theorem}[section]
\newtheorem{cor}{Corollary}[section]
\newtheorem{lem}{Lemma}[section]

\newtheorem{exa}{Example}[section]

\def\F{{\mathbb F}}

\newcommand\vu{{\mathbf{u}}}
\newcommand\vv{{\mathbf{v}}}
\newcommand\vM{{\mathbf{M}}}
\newcommand\vd{{\mathsf{d}}}
\newcommand\vD{{\mathsf{D}}}

\newcommand{\floor}[1]{{\left\lfloor #1 \right\rfloor}}

\begin{document}

\title{Multiply Constant-Weight Codes and the Reliability of Loop Physically Unclonable Functions}
\author{
	Yeow Meng Chee,~\IEEEmembership{Senior Member, IEEE},  
        	Zouha Cherif, 
        	Jean-Luc Danger, 
        	Sylvain Guilley, 
        	Han Mao Kiah,
        	Jon-Lark Kim, 
        	Patrick Sol\'{e},
	 Xiande Zhang
	 
	 \thanks{Research of Y. M. Chee, H. M. Kiah and X. Zhang is supported in part by 
Singapore National Research Foundation under Research Grant NRF-CRP2-2007-03.
Research of Z. Ch\'erif, S. Guilley and J.-L. Danger is supported by Orange Labs and 
the ENIAC European project 2010-1 ``TOISE'' (Trusted Computing for European Embedded Systems). 
Research of J.-L. Kim is supported by Basic Research Programme through the National Research Foundation of Korea (NRF)
funded by Ministry of Education (NRF-2013R1A1A2005172) and 
by the Sogang University Research Grant of 201210058.01.
This paper was presented in part at the IEEE International Symposium on Information Theory, 2013.}
\thanks{Y. M. Chee, H. M. Kiah and X. Zhang 
are with the Division~of~Mathematical Sciences,
  School~of~Physical~and~Mathematical~Sciences,
  Nanyang~Technological~University, 21~Nanyang~Link, Singapore~637371,
  Singapore (emails:\{YMChee, HMKiah, XiandeZhang\}@ntu.edu.sg).}
\thanks{Z. Cherif, J.-L. Danger, S. Guilley and P. Sol\'e are with 
Institut MINES-TELECOM, TELECOM ParisTech, CNRS LTCI, 46 rue Barrault, 75\,634 Paris, France
(email:\{zouha.cherif,jean-luc.danger,sylvain.guilley,patrick.sole\}@telecom-paristech.fr).}
\thanks{J.-L. Kim is with
Department of Mathematics, Sogang University, Seoul 121-742, {South Korea}
(emal:jlkim@sogang.ac.kr).}
\thanks{Z. Cherif is also with 
Universit\'e de Lyon, CNRS, UMR5516, Laboratoire Hubert Curien 42\,000, Saint-\'Etienne, France.}
\thanks{J.-L. Danger and S. Guilley are also with 
Secure-IC S.A.S., 80 avenue des Buttes de Co\"esmes, 35\,700 Rennes, {France}.}
\thanks{P. Sol\'e is also with 
King Abdulaziz University, Department of Mathematics, Jeddah, {Saudi Arabia}.}
}

%
%
%
%
%
%
%
%
%

\date{}

\maketitle
\begin{abstract}
We introduce the class of multiply constant-weight codes
to improve the reliability of certain physically unclonable function (PUF) response.
We extend classical coding methods to construct multiply constant-weight codes 
from known $q$-ary and constant-weight codes.
Analogues of Johnson bounds are derived and 
are shown to be asymptotically tight to a constant factor under certain conditions.
We also examine the rates of the multiply constant-weight codes and interestingly, 
demonstrate that these rates are the same as those of constant-weight codes of suitable parameters.
Asymptotic analysis of our code constructions is provided.

\end{abstract}

\pagebreak

\section{Introduction}
\label{sec_into}

Physically unclonable functions (PUFs) introduced by Pappu {\em et al.}\cite{Pappuetal:2002}
provide innovative low-cost authentication methods that are derived from complex physical characteristics of electronic devices.
Recently, PUFs have become an attractive option to provide security in low cost devices 
such as RFIDs and smart cards \cite{Pappuetal:2002, Gassendetal:2002, SuhDevadas:2007,Cherifetal:2012}.
Reliability and implementation considerations on programmable circuits for 
the design of Loop PUFs \cite{Cherifetal:2012} lead to the 
investigation of a new class of codes called {\em multiply constant-weight codes} (MCWC).

In an MCWC, 
each codeword is a binary word of length $mn$ 
which is partitioned into $m$ equal parts and has weight exactly $w$ in each part \cite{Cherif:2013}.
This definition therefore generalizes the class of {\em constant-weight codes} (where $m=1$) and
a subclass of {\em doubly constant-weight codes}, introduced by Johnson\cite{Johnson:1972} and Levenshte{\u\i}n 
\cite{Levenshtein:1971} (where $m=2$).

In this paper, we consider upper and lower bounds for the possible sizes 
of MCWCs. 
Our constructions make use of both classical concatenation techniques \cite{Forney:1966, Dumer:1998}
and a method due to Zinoviev (for constant-weight codes)\cite{Zinoviev:1983}, 
that was later independently given by Etzion (for doubly constant-weight codes) \cite{Etzion:2007}.
A construction technique using resolvable designs is also examined.
For upper bounds, we extend the techniques of Johnson \cite{Johnson:1972} and 
exhibit that these bounds are asymptotically tight to a constant factor, provided $m$, $w$ and $d$ are fixed.
We also examine the rates of the MCWCs 
 and interestingly, demonstrate that these rates are the same as those of constant-weight codes of length $mn$ and weight $mw$.

We remark that if the codewords in an MCWC 
are regarded as $m$ by $n$ arrays,
then an MCWC 
can be regarded as a code over binary matrices, where
each matrix has constant row weight $w$. 
These codes were studied by Chee {\em et al.} \cite{Cheeetal:2013c} in an application for power line communications.
The relevance of MCWCs for the latter context is an area for future research.

The rest of this article is structured as follows. 
Section~\ref{sec_notations} collects the necessary definitions and notation, and
Section~\ref{sec_motiv} examines an application of MCWCs in the field of PUFs.
Section~\ref{sec_lb} deals with constructions and attached lower bounds, while
Section~\ref{sec_ub} contains the upper bounds. 
Section~\ref{sec_asymp} studies asymptotic versions of the bounds of Section~\ref{sec_lb} and Section \ref{sec_ub}.
Some of our results were initially reported in \cite{Cherif:2013} and 
the present paper contains many new results and generalizations.

\section{Definitions and Notation}
\label{sec_notations}

Let $\cal X$ be a set of $q$ symbols.
A $q$-ary {\em code} $C$ of length $n$ over the alphabet $\cal X$ is a subset of ${\cal X}^n$. 
Elements of $C$ are called {\em codewords}.
Endow the space ${\cal X}^n$ with the {\em Hamming distance} metric. 
A code $C$ 
is said to have {\em distance $d$} 
if the (Hamming) distance between any two distinct codewords of $C$ is at least $d$. 
A $q$-ary code of length $n$ and distance $d$ is called an $(n,d)_q$ code. 

When $q=2$, we assume ${\cal X}=\F_2$. An $(n,d)_2$ code is simply called an $(n,d)$ code. 
The (Hamming) {\em weight}  of a codeword $\vu\in{\cal X}^n$ is given by the number of nonzero coordinates in $\vu$.
Fix $m, n_1, n_2, \ldots, n_m$ to be positive integers and let $N=n_1+n_2+\cdots+n_m$.
An $(N,d)_2$ code is said to be of {\em multiply constant-weight} and denoted by  MCWC$(w_1,n_1;w_2,n_2; \ldots ; w_m, n_m;d)$,
if each codeword has weight $w_1$ in the first $n_1$ coordinates, weight $w_2$ in the next $n_2$ coordinates, and so on and so forth.
When $m=1$, an MCWC$(n,w;d)$ is a {\em constant-weight code}, denoted by CWC$(n,d,w)$;
when $m=2$, an MCWC$(w_1,n_1; w_2,n_2; d)$ is a {\em doubly constant-weight code}.

When $w_1=w_2=\cdots=w_m=w$ and $n_1=n_2=\cdots=n_m=n$, 
we simply denote this multiply constant-weight code of length $N=mn$ by MCWC$(m,n,d,w)$. 
Unless specified otherwise, a multiply constant code refers to an MCWC$(m,n,d,w)$
in this paper.

The largest size of an $(n,d)_q$ code is denoted by $A_q(n,d)$. 
When $q=2$, this size is simply denoted by $A(n,d)$. 
The largest of size of an MCWC$(w_1,n_1;w_2,n_2; \ldots ; w_m, n_m;d)$ is given by $T(w_1,n_1;w_2,n_2; \ldots ; w_m, n_m;d)$;
 the largest of size of an MCWC$(m,n,d,w)$ is given by $M(m,n,d,w)$;
 and the largest of size of a CWC$(n,d,w)$ is given by $A(n,d,w)$.

 In this paper, we are mainly interested in determining $M(m,n,d,w)$. 
 Observe that by definition,  
 \begin{align*}
 M(1,n,d,w)	&= A(n,d,w), \\
M(2,n,d,w)   	&= T(w,n;w,n;d).
 \end{align*}
Moreover, the functions $A(n,d,w)$ and $T(w,n;w,n;d)$ have been well studied 
(see for example, \cite{Johnson:1972, Brouweretal:1990, Agrelletal:2000, Smithetal:2006, Etzion:2007}).
Online tables of the lower bounds for $A(n,d,w)$ can be found at \cite{Brouwer} 
while upper bounds for $A(n,d,w)$ and $T(w,n;w,n;d)$ can be found at \cite{Agrell}. 

In this paper, we are mainly interested in building multiply constant-weight codes 
from known $q$-ary codes and constant-weight codes.
One such class of codes is the class of binary {\em linear codes}. 
A binary linear code of length $n$, dimension $k$ and distance $d$ is called a linear $[n,k,d]$ code
and we denote the largest quantity $2^k$ of a binary linear $[n, k, d]$ code by $B(n,d)$.

Unfortunately, an MCWC 
cannot be linear and hence, we look at possible generalization of linearity. 
A possible generalization given by the notion of systematic codes.
A code of size $2^k$ is said to be {\em systematic}
if there is a set $I$ of $k$ coordinates such that the code 
when restricted to the coordinate set $I$ is exactly $\F_2^k$.
The largest sizes of a systematic $(n,d)$ code and CWC$(n,d,w)$ are
denoted by $S(n,d)=2^{s(n,d)}$ and $S(n,d,w)=2^{s(n,d,w)}$ respectively.
We remark that systematic constant-weight codes have been studied in 
\cite{BoinckvanTilborg:1990, Lin:1993}.

Finally, as mentioned in the introduction, a codeword in an MCWC$(m,n,d,w)$ 
can be regarded as a binary $m$ by $n$ matrix with constant row weight $w$.
Throughout the rest of this paper, we shall regard a codeword in an MCWC 
as either a word of length $mn$ or an $m$ by $n$ matrix.

%

\section{Application to Loop PUFs}
\label{sec_motiv}

The need of an MCWC  arises from the generation of some type of PUFs in trusted electronic circuits.
In this section, we demonstrate the relevance of MCWC in the implementation of Loop PUF 
on Field Programmable Gate Array (FPGA) and 
in enhancing the reliability of PUF response.
First, we present the principle behind Loop PUF.

\subsection{Loop PUF Principle}
In general, the PUF provides a unique signature to a device without 
the need for the user to program an internal memory~\cite{Pappuetal:2002}.
This signature allows the user to build lightweight authentication protocols or 
even protect a master key in cryptographic implementations.
Such a key can be used for standard cryptographic protocols, or for internal cryptography ({\em e.g.}, memory encryption).
Essentially, the PUF takes advantage of technological process variations to differentiate between two devices.
For instance, consider two delay lines with the same structure. 
In theory, the propagation time is the same for both two delay lines.
However, actual measurements of the propagation time differ between the delay lines 
due to imbalances between the physical elements.
Furthermore, as these measurements cannot be predicted accurately, 
they are well suited for cryptographic purposes.

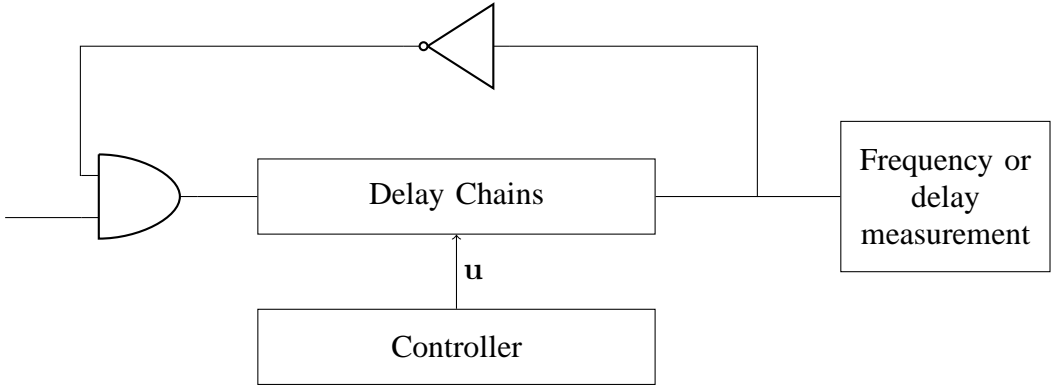
\begin{figure}
\centering
\begin{tikzpicture}

\node[style={draw,rectangle, text width=5cm, minimum height=1cm, text badly centered}, name=delay] at (5,2) {Delay Chains} ;
\node[style={draw,rectangle, text width=5cm, minimum height=1cm, text badly centered}, name=control] at (5,0) {Controller} ;
\node[style={draw, rectangle, text width=2.5cm, minimum height=2cm, text badly centered}, name=output] at (11.5,2) {Frequency or delay\\ measurement};
\draw
 (1.4,2) node[and port]  (myand) {} 
 (5,4) node[not port, rotate=180] (inverter){}
 (-1,1.72)--(myand.in 2)
(myand.out)--(delay)--(9,2)--(9,4) --(inverter.in)
(inverter.out)--(0,4)|-(myand.in 1)
;
\draw (9,2)--(output);
\draw [->] (control) -- node {~~~$\mathbf u$} (delay);

\end{tikzpicture}

\caption{Loop PUF structure}
\label{fig:loop}

\end{figure}

Here, we consider the Loop PUF~\cite{Cherifetal:2012} that is a set of $n$ identical delay lines laid out on a programmable circuit. 
The delay lines form a loop that oscillates as a single ring oscillator when closed by an inverter (see Figure \ref{fig:loop})
and this setup enhances the accuracy of delay measurements.
Furthermore, each delay line is a series of $m$ delay elements and  
the delay of the $i$th element of the $j$th line is controlled by the $(i,j)$-th bit of some control word $\vu$ of length $mn$.
Hence, corresponding to a control word $\vu$, we have a delay measurement, denoted by $\mathsf D(\vu)$.

For expository purposes, we illustrate how a general binary code can be used in conjunction with the Loop PUF \cite{Cherifetal:2012} to 
generate a set of Challenge-Response pairs for authentication purposes.
For other cryptographic applications, we refer the interested reader 
to \cite{Pappuetal:2002, Gassendetal:2002, SuhDevadas:2007,Cherifetal:2012}.

Given a binary code $C$ of length $mn$, the set of Challenge-Response pairs is given by
\begin{equation*}
\left\{ \Big((\vu,\vv), \mathsf{sign} (\mathsf{D}(\vu)-\mathsf{D}(\vv)) \Big):\vu\ne\vv, \vu,\vv \in C\right\} \enspace.
\end{equation*}
In other words, each challenge is an ordered pair of distinct codewords $(\vu, \vv)$ from the binary code
and the corresponding response is the sign of the delay difference between the pair of codewords%
\footnote{Here, we consider a response that consists of {\em one} bit. 
A different control strategy can be used to extract a response with more bits
and this is described in \cite{Cherifetal:2012}.}.

For the set of Challenge-Response pairs to be used for authentication, it is important that 
we are unable to infer the sign of the delay difference with only knowledge of $\vu$ and $\vv$.
To achieve this {\em unpredictability} of response, we show that $C$ needs to be an MCWC in Section \ref{sub-mcwc}.
On the other hand, it is also important that the measured response (or the sign of delay difference) remains the same 
despite environmental noise. 
The {\em reliability} of response is then shown to be associated with the minimum distance of the code $C$ in Section \ref{sub-reliability}.
Therefore, MCWCs are needed to satisfy both requirements of unpredictability and reliability.

\subsection{MCWC to Achieve Unpredictability on FGPAs}\label{sub-mcwc}

Programmable circuits, like FPGAs, have a hierarchical layout.
It is thus convenient to organize the PUF with two levels, namely with a structure of $n$ clusters of $m$ cells each%
\footnote{Logic Array Block (LABs) for ALTERA and Configurable Logic Blocks (CLBs) for XILINX}.
For this technology, it is rather easy to copy / paste exactly the logic of one cluster to generate all of them, in an indistinguishable manner (logically, not physically).
Thus the Loop PUF can be easily constructed from a set of $n$ clusters of $m$ cells just by replicating the base cluster of $m$ cells.
As the routing inside a cluster between the $m$ elements is not constrained, the PUF designer can easily port this structure to any FPGA family.  

Now, let us consider an MCWC$(n,w_1;n,w_2;\cdots;n,w_m;d)$ and choose a control word  $\vu= (u_{11},u_{12},\ldots,u_{1n},u_{21},u_{22},\ldots,u_{2n},\ldots, u_{m1},u_{m2},\ldots,u_{mn})$.
Let $\vd_{ij}(u_{ij})$ be the resulting delay of the $i$th delay element in the $j$th line and 
hence, the total measured delay $\vD(\vu)$ due to $\vu$ is given by 
$\sum_{i=1}^m\sum_{j=1}^n \vd_{ij}(u_{ij})$.

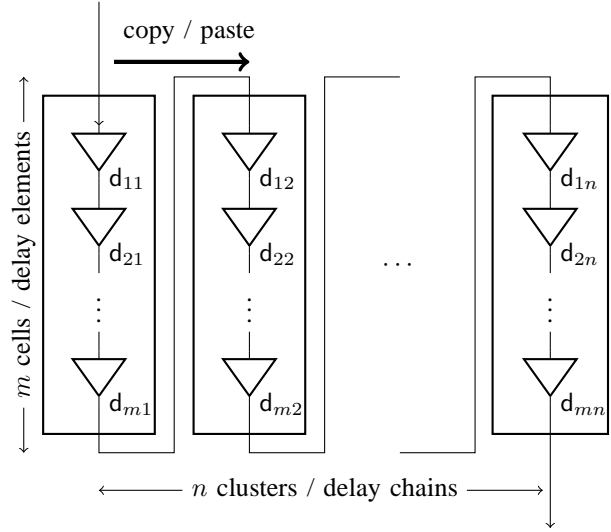
\begin{figure}
\centering
\small
\begin{tikzpicture} 

\node[style={draw,thick, rectangle, text width=1.25cm, minimum height=4.5cm}, name=chain1] at (0,2.5) {} ;
\node[style={draw,thick, rectangle, text width=1.25cm, minimum height=4.5cm}, name=chain2] at (2,2.5) {} ;
\node[style={draw,thick, rectangle, text width=1.3cm, minimum height=4.5cm}, name=chainn] at (6,2.5) {} ;

\node[name=vdots1] at (0,2) {$\vdots$};
\node[name=vdots2] at (2,2) {$\vdots$};
\node[name=vdotsn] at (6,2) {$\vdots$};
\node[] at (4,2.5) {$\cdots$};

\node[name=labeln] at (3,-0.5) {$n$ clusters / delay chains};
\draw[->] (labeln)--(5.9,-0.5);
\draw[->, bend right] (labeln)--(0,-0.5);

\node[name=labeln,rotate=90] at (-1,2.5) {$m$ cells / delay elements};
\draw[->] (labeln)--(-1,0);
\draw[->] (labeln)--(-1,5);

\node[] at (1.2,5.6){copy / paste};
\draw[->,ultra thick] (0.2,5.2)--(2,5.2);

\draw
 (0,4) node[buffer, rotate=-90,scale=0.5]  (buff11) {}
 (0,3) node[buffer, rotate=-90,scale=0.5]  (buff21) {} 
 (0,1) node[buffer, rotate=-90,scale=0.5]  (buffm1) {} 
 
 (2,4) node[buffer, rotate=-90,scale=0.5]  (buff12) {}
 (2,3) node[buffer, rotate=-90,scale=0.5]  (buff22) {} 
 (2,1) node[buffer, rotate=-90,scale=0.5]  (buffm2) {} 
 
 (6,4) node[buffer, rotate=-90,scale=0.5]  (buff1n) {}
 (6,3) node[buffer, rotate=-90,scale=0.5]  (buff2n) {} 
 (6,1) node[buffer, rotate=-90,scale=0.5]  (buffmn) {} 

 (buff11.out)--(buff21.in) (buff21.out)--(vdots1)--(buffm1.in) (buffm1.out)--(0,0)--(1,0)--(1,5)-|(buff12.in)
  (buff12.out)--(buff22.in) (buff22.out)--(vdots2)--(buffm2.in) (buffm2.out)--(2,0)--(3,0)--(3,5)--(4,5)
 (4,0)--(5,0)--(5,5)-|(buff1n.in)
  (buff1n.out)--(buff2n.in) (buff2n.out)--(vdotsn)--(buffmn.in) 
;
\draw[->](0,6)--(buff11.in);
 \draw[->](buffmn.out)--(6,-1);

 \node[right=0.01cm] at (buff11.out) {$\mathsf d_{11}$};
 \node[right=0.01cm] at (buff21.out) {$\mathsf d_{21}$};
 \node[right=0.01cm] at (buffm1.out) {$\mathsf d_{m1}$};

\node[right=0.01cm] at (buff12.out) {$\mathsf d_{12}$};
 \node[right=0.01cm] at (buff22.out) {$\mathsf d_{22}$};
 \node[right=0.01cm] at (buffm2.out) {$\mathsf d_{m2}$};

\node[right=0.01cm] at (buff1n.out) {$\mathsf d_{1n}$};
 \node[right=0.01cm] at (buff2n.out) {$\mathsf d_{2n}$};
 \node[right=0.01cm] at (buffmn.out) {$\mathsf d_{mn}$};

\end{tikzpicture}

\caption{Delay chain layout}
\label{fig:delay}

\end{figure}

Ideally, $\vd_{ij}(u_{ij})=\mu+\epsilon_{ij}(u_{ij})$, where $\epsilon_{ij}$ is a small timing variation on the $j$th delay element on the $i$th line caused by technological dispersion and $\mu$ is the average delay that is independent of the position on the circuit. 
However, the latter is not true due to manufacturing constraints.
In particular, a designer has no control about the routing within an FPGA cluster and 
hence, it is hardly possible to get balanced delay elements within a cluster.
But fortunately due to copy / paste operation, the internal routing of a cluster can be faithfully reproduced 
from one cluster to another (see Figure \ref{fig:delay}).

In other words, we have
\begin{equation*}
\vd_{ij}(u_{ij})=\mu_i(u_{ij})+\epsilon_{ij}(u_{ij}),
\end{equation*}
where $\epsilon_{ij}$ is a small timing variation and $\mu_i$ is the average delay dependent on the controlled bit 
and the position of the delay element. We compute the total delay due to $\vu$, and we have

\begin{align}
\vD(\vu)&
= \sum_{i=1}^m\sum_{j=1}^n \vd_{ij}(u_{ij}) \notag \\
&= \sum_{i=1}^m\sum_{j=1}^n \mu_i(u_{ij})+\epsilon_{ij}(u_{ij}) \notag\\
&= \left(\sum_{i=1}^m (n-w_i)\mu_i(0)+w_i\mu_i(1)\right)+\left(\sum_{i=1}^m\sum_{j=1}^n \epsilon_{ij}(u_{ij})\right) .\label{eq:delay}
\end{align}
\noindent The last equality follows from the fact that $\vu$ belongs to an MCWC$(n,w_1;n,w_2;\cdots;n,w_m;d)$. 
Furthermore, we observe that all codewords from the MCWC have the same expected response.
Therefore, the delay difference between any pair of control words from the MCWC has expectation zero
and the sign of the difference is dependent only on $\epsilon_{ij}$'s. 
In other words, the response depends entirely on the unpredictable physical characteristics of the individual delay elements.

\subsection{Hamming Distance to Improve the PUF Reliability}\label{sub-reliability}
The PUF response is very sensitive to environmental noise as the $\epsilon_{ij}$ can be very low in comparison to the delays.
Hence it is necessary to choose pairs of control words which offer the largest possible difference between their delays.

From \eqref{eq:delay}, we see that
\begin{equation*}
\mathsf{D}(\vu)-\mathsf{D}(\vv)=\sum_{i=1}^m\sum_{j=1}^n \epsilon_{ij}(u_{ij})-\epsilon_{ij}(v_{ij})
=\sum_{u_{ij}\ne v_{ij}} \ \epsilon_{ij}(u_{ij})-\epsilon_{ij}(v_{ij}) \enspace.
\end{equation*}
\noindent Therefore, the greater the Hamming distance between $\vu$ and $\vv$, 
the greater the delay difference $\mathsf{D}(\vu)-\mathsf{D}(\vv)$.
Hence, by choosing a code of high distance, we improve the reliability of the PUF response. 

The arguments in this section demonstrate the relevance of MCWC in the design of reliable Loop PUF. 
In the remaining of the paper, we examine the possible lower and upper bounds for optimal MCWCs,
focusing our attention to the case where $w_1=w_2=\cdots=w_m=w$.

\section{Lower Bounds}
\label{sec_lb}
\subsection{Coding Constructions}
\label{ssec_cc}
In this section, we study constructions of MCWCs 
using known unrestricted codes.
Our first construction is based on {\em concatenation}.

\begin{prop}
\label{lb1} 
Let $q\le A(n,d_1,w).$  
We have $$M(m,n,d_1d_2,w)\ge A_q(m,d_2). $$
\end{prop}

\begin{IEEEproof}
Consider a concatenation scheme \cite{Forney:1966, Dumer:1998} where 
the {\em outer code} $C$ is an $(m,d_2)_q$ code of size $A_q(m,d_2)$ over $\cal X$
and the {\em inner code} $D$ is a CWC$(n,d_1,w)$ of size $q$.
Let $\phi:\mathcal X \to D$ be an injective map.
For each codeword $\vu=(\vu_1,\vu_2, \ldots, \vu_m)$ in $C$, 
we construct the binary codeword $(\phi(\vu_1),\phi(\vu_2),\ldots,\phi(\vu_m))$.
Then the resulting code is an MCWC$(m,n,d_1d_2,w)$ of size $A_q(m,d_2)$.
\end{IEEEproof}

A special case of concatenation is the {\em product code construction}.
Recall that if $C$ and $D$ are two binary linear codes 
then their product $C\otimes  D$ is the code of length $nm$ consisting of $m$ by $n$ arrays 
whose rows belong to $C$ and columns belong to $D$.
If $C$ and $D$ are linear $[m,k,d]$ and $[n,l,e]$ codes, 
then the code $C\otimes D$ has parameters $[nm,kl,de]$ \cite[Lemma 2.8]{Dumer:1998}. 

We generalize this construction by relaxing certain requirements.
In particular, we require only the rows of our arrays to be in $C$, 
while not all the columns need to be in $D$.
Formally, consider a systematic CWC$(n,d_1,w)$ $C$ of size $2^{k_1}$ and
a systematic $(m,d_2)$ code $D$ of size $2^{k_2}$.

Given a binary $k_2$ by $k_1$ matrix $\vM$, 
we replace each column of length $k_2$ of $\vM$ with its corresponding codeword in $D$
and we obtain a binary $m$ by $k_1$ matrix $\vM'$. 
Next replace each row of length $k_1$ of $\vM'$ with its corresponding codeword in $C$.
This results in a binary $m$ by $n$ matrix with constant row weight $w$.
In particular, each row of the matrix belongs to the constant-weight code $C$ 
while the first $k_1$ columns belong to the code $D$. 
Hence, the collection of all $2^{k_1k_2}$ matrices from this construction results 
in an MCWC$(m,n,d_1d_2,w)$. We call this construction a {\em pseudo-product code construction}.


We remark that as with product construction, the pseudo-product code construction is a special case of concatenation.
In addition, the pseudo-product construction coincides with the construction
given by Amrani \cite[Definition 1]{Amrani:2007} in another context.
The following proposition follows immediately from the pseudo-product code construction.


\begin{prop}
\label{sys}  
We have 
$$M(m,n,d_1 d_2,w)\ge 2^{s(n,d_1,w)s(m,d_2)}\ge B(m,d_2)^{s(n,d_1,w)}. $$
\end{prop}

\begin{exa}
Consider the following systematic constant-weight code $\{ 0011,0101,1010,1111\}$
of distance two. 
Taking its pseudo-product with a binary linear $[6,2,4]$ code yields 
a lower bound of $2^{2\cdot 2}=16$ on $M(6,4,8,2).$ 
\end{exa}


We give a simple but robust construction technique for systematic constant-weight codes due to B\"oinck and van Tilborg.


\begin{prop}[B\"oinck and van Tilborg{\cite[Construction 4.1]{BoinckvanTilborg:1990}}]\label{simple}  
We have $$S(2n,2d,n)\ge S(n,d)\ge B(n,d) . $$
\end{prop}

\begin{IEEEproof}
Let $C$ be a systematic code of size $S(n,d).$ Construct a constant-weight code by the rule
$$D=\{ (x,\overline{x}) \vert \, x \in C \},$$ where the bar denotes complementation. 
The code $D$ hence has twice the distance of $C$ and is systematic because $C$ is.
 \end{IEEEproof}

\begin{exa}
Observe that $B(2^{m-1},2^{m-2})=2^m$ follows from the Plotkin bound and 
the Reed Muller code $RM(1,m-1)$ \cite[Chapter 13]{MacWilliamsSloane:1977}.
Proposition \ref{simple} therefore yields $S(2^m,2^{m-1},2^{m-1})\ge 2^{m}$.
\end{exa}

We extend the code construction in Proposition \ref{simple} by appending each codeword 
with a codeword from a suitable constant-weight code.
\begin{prop}\label{extend} If $2^k\le A(n,d,w)$ we have $$s(n+2k,d+2,w+k)\ge k . $$
\end{prop}

\begin{IEEEproof}
Let $C$ be a constant-weight code of size $A(n,d,w).$ 
Let $\phi:\F_2^k \to C$ be an injective map.
Let
$$D=\{ (x,\overline{x}, \phi(x)) \vert \, x \in \F_2^k \},$$ where the bar denotes complementation. The code $D$ is systematic with information set the first $k$ coordinates and has the required parameters.
 \end{IEEEproof}



The next construction generalizes a construction by Zinoviev \cite{Zinoviev:1983} 
(see also \cite{KautzSingleton:1964})
and by Etzion \cite[Theorem 16]{Etzion:2007}
to construct multiply constant-weight codes from $q$-ary codes.

\begin{prop}\label{qary} We have 
$$M(m,qw,2d,w)\ge A_q(mw,d). $$
\end{prop}

\begin{IEEEproof}
%
%
Consider an $(mw,d)_q$ code of size $A_q(mw,d)$ over the alphabet $\cal X$.
We extend each word of length $mw$ to a word of length $qmw$
by replacing each symbol with a binary word of length $q$.
Specifically, replace each symbol in the codeword with 
the following characteristic function $\phi:\mathcal X\to \{0,1\}^{\mathcal X}$,

\begin{equation*}
\phi(x)_{y}=
\begin{cases}
1, & \mbox{if } x=y,\\
0, & \mbox{otherwise}.
\end{cases}
\end{equation*}

We check that the new binary word of length $qmw$ comprises $m$ parts 
each of weight $w$.

It remains to check that the distance.
Observe that for any pair of distinct symbols $x,y\in \mathcal X$, the distance between $\phi(x)$ and $\phi(y)$ is two.
Hence, since the distance between two $q$-ary codewords is at least $d$,
the distance between the corresponding binary codewords is at least $2d$.
\end{IEEEproof}

When $q$ is a prime power and  $q\ge mw-1$, there exists a $q$-ary Reed Solomon code of length $mw$ and distance $d$.
Hence, $A_q(mw,d)=q^{mw-d+1}$ and the following corollary is immediate.

\begin{cor}
If $q$ is a prime power and  $q\ge mw-1$,
then 
\begin{equation*}
M(m,qw,2d,w)\ge q^{mw-d+1}. 
\end{equation*}
\end{cor}

On the other hand, when $w=1$, we observe that 
we are able to reverse the construction so
as to construct an $n$-ary codeword of length $m$
from an $m$ by $n$ matrix with constant row weight one.
Hence, the following corollary is immediate.

\begin{cor}\label{cor:rs}
We have
\begin{equation*}
M(m,n,2d,1)=A_n(m,d).
\end{equation*}
\end{cor}

\subsection{Designs Constructions}
\label{ssec_dc}

Here, we consider a construction from designs, in particular, {resolvable $t$-designs}.

A $t$-$(v,k,1)$ design, or $t$-design, is a pair $(X,\mathcal B)$ such that
 $|X| = v$ and $\cal B$ is a collection of $k$-subsets of $X$, called {\em blocks}, 
 with the property that every $t$-subset of $X$ is contained in exactly one block. 
 A $t$-design $(X,\mathcal B)$ is {\em resolvable} if the blocks in $\cal B$ 
 can be partitioned into {\em parallel classes}, 
 each of which is a partition of $X$.
 
 Suppose $(X,\mathcal B)$ is a resolvable $t$-$(v,k,1)$ design
 with $M= k\binom{v}{t}/v\binom{k}{t}$ parallel classes.
Let $m=v/k$ and $n=v$. For each parallel class, we construct a binary $m$ by $n$ matrix,
 where the support of each row is given by a corresponding block. 
 Hence, we form a binary $m$ by $n$ matrix with constant row weight $k$.
 Since every pair of blocks intersect at most in $t-1$ places, 
 the distance between every pair of binary matrices is at least $2m(k-t+1)$.
 Hence, we obtain an MCWC$(m,n,2m(k-t+1),k)$ of size $M$.
 We summarize the construction in the following proposition.
 
 \begin{prop}\label{prop:design}
 Suppose there exists a resolvable $t$-$(v,k,1)$ design.
 Then 
 \begin{equation*}
 M\left( \frac vk, n, 2(k-t+1)\frac vk, k\right)\ge \frac{k\binom{v}{t}}{v\binom{k}{t}}.
 \end{equation*}
 \end{prop}

Existence results for resolvable $2$-$(v,k,1)$ design are surveyed by Abel {\em et al.} \cite[Table 7.35]{Abeletal:2007b}.
When $t\ge 3$, existence results are given by Laue \cite{Laue:2004} (see also \cite{Krameretal:1980, vanTrung:2000, vanTrung:2001}).


\section{Upper Bounds}
\label{sec_ub}
Trivially, an MCWC$(m,n,d,w)$ is a CWC$(mn,d,mw)$. 
Hence, we have our first upper bound.

\begin{prop} \label{triv}We have $$M(m,n,d,w)\le A(nm,d,mw).$$
\end{prop}

Next, we extend the techniques of Johnson \cite{Johnson:1972} to obtain 
the following recursive bounds on $T(w_1,n_1;w_2,n_2;\ldots; w_m, n_m;d)$.
Let $1\le i\le m$.
\begin{align}
T(w_1,n_1;w_2,n_2;\ldots; w_m, n_m;d) &
\le \floor{\frac{n_i}{w_i}T(w_1,n_1;\ldots;w_i-1,n_i-1;\ldots; w_m,n_m;d)}, \label{eq:john1}\\
T(w_1,n_1;w_2,n_2;\ldots; w_m, n_m;d) &
\le \floor{\frac{n_i}{n_i-w_i}T(w_1,n_1;\ldots;w_i,n_i-1;\ldots; w_m,n_m;d)},\\
T(w_1,n_1;w_2,n_2;\ldots; w_m, n_m;d) &
\le \floor{\frac{u}{w_1^2/n_1+w_2^2/n_2+\cdots+w_m^2/n_m-\lambda}}, &
\end{align}
\noindent where $d=2u$ and $\lambda=w_1+w_2+\cdots+w_m-u$.
Since $M(m,n,d,w)=T(w,n;w,n; \ldots; w,n; d)$ and 
applying the recursive bounds $m$ times, we obtain the following recursive upper bounds.

\begin{prop}\label{recur-bdd}
We have
\begin{align}
M(m,n,d,w) & \le \floor{\frac{n^m}{w^m} M(m, n-1, d, w-1)}\label{eq:mcwc1}\\
M(m,n,d,w) & \le \floor{\frac{n^m}{(n-w)^m} M(m, n-1, d, w)}\\
M(m,n,d,w) &\le \floor{\frac{d/2}{mw^2/n - (mw-d/2)}}.
\end{align}
\end{prop}

%
%
%
%
%
%
%
%
%
%
%
%
%
%

Suppose $s=mw-d/2+1\le m$. Applying \eqref{eq:john1} for $s$ iterations,
we have $M(m,n,d,w)\le \frac{n^s}{w^s}T(w-1,n-1;\ldots;w-1,n-1;w,n;\ldots;w,n;d)$ and 
$T(w-1,n-1;\ldots;w-1,n-1;w,n;\ldots,w,n;d)$ is trivially one.
Hence, we obtain the next upper bound.

\begin{prop}\label{prop:singleton-like}
If $mw-d/2+1\le m$, then 
\begin{equation}\label{eq:mcwc2}
M(m,n,d,w)\le \left(\frac nw\right)^{mw-d/2+1}.
\end{equation}
\end{prop}



We remark that when $w=1$, Proposition \ref{prop:singleton-like}
reduces to the classical Singleton bound.

\vskip 5pt

Given $m$, $d$, $w$, let $i$ be the smallest integer 
such that $m(w-i)-d/2+1\le m$. 
Then $i$ iterative applications of \eqref{eq:mcwc1}, followed
by an application of \eqref{eq:mcwc2}, yields the following corollary.

\begin{cor}\label{cor:johnson}
Given $m$, $d$, $w$, let $i$ be the smallest integer 
such that $m(w-i)-d/2+1\le m$ and $t=m(w-i)-d/2+1$. 
Then we have
{
\begin{align*}
M(m,n,d,w) 
&\le \floor{\frac{n^m}{w^m}\floor{\frac{(n-1)^m}{(w-1)^m} \cdots \floor{\frac{(n-i+1)^m}{(w-i+1)^m}\floor{\frac {(n-i)^t}{(w-i)^t}}}\cdots}}\\
&\le \frac{n^{mw-d/2+1}}{(w-i)^{mw-d/2+1}}. 
\end{align*}
}
\end{cor}

When the $m$, $d$ and $w$ are fixed, we establish
tightness of the bound given by Corollary \ref{cor:johnson}.

\begin{cor}\label{cor:tightness}
Fix $m$, $d$ and $w$. Let $s=mw-d/2+1$ and 
$i$ be the smallest integer 
such that $m(w-i)-d/2+1\le m$.

Consider $M(m,n,d,w)$ as a function of $n$. We have
\begin{equation}\label{eq:Mmndw}
1\le \limsup_{n\to\infty} \frac{M(m,n,d,w)}{n^s/w^s} \le \frac{w^s}{(w-i)^s}.
\end{equation} 
In addition, when $s\le m$, $n/w\ge mw-1$ and $n/w$ is a prime power,
we have 
\begin{equation*}
M(m,n,d,w)=\frac {n^s}{w^s}.
\end{equation*} 
\end{cor}

\begin{IEEEproof}
When  $n/w\ge mw-1$ and $n/w$ is a prime power,
Corollary \ref{cor:rs} establishes that 
\begin{equation*}
\limsup_{n\to\infty} \frac{M(m,n,d,w)}{n^s/w^s} \ge 1.
\end{equation*} 
Then Corollary \ref{cor:johnson} establishes \eqref{eq:Mmndw}.

If in addition, when $s\le m$, Proposition \ref{prop:singleton-like} with Corollary \ref{cor:rs} 
establishes that $M(m,n,d,w)=(n/w)^s$.
\end{IEEEproof}

\section{Asymptotics}
\label{sec_asymp}

In this section, we consider the asymptotic rate of $M(m,n,d,w)$ when $m$ is large, 
$n$ is a function of $m$, 
$d=\floor{\delta nm}$ and $w=\floor{\omega n}$ for $0<\delta,\omega<1$.
Specifically, we determine the value $\mu(\delta,\omega)$, where
\begin{equation*}
\mu(\delta, \omega):=\limsup_{m \to\infty} \frac{\log_2 M(m,n,\floor{\delta mn},\floor{\omega n})}{mn}.
\end{equation*}

In the following discussion, we make use of the following better known exponents.
\begin{align*}
\alpha_q(\delta) &:=\limsup_{n \to\infty} \frac{\log_q A(n,\floor{\delta n})}{n},\\
\alpha(\delta) &:=\limsup_{n \to\infty} \frac{\log_2 A(n,\floor{\delta n})}{n},\\
\alpha(\delta,\omega) &:=\limsup_{n \to\infty} \frac{\log_2 A(n,\floor{\delta n},\floor{\omega n})}{n},\\
\sigma(\delta) &:=\limsup_{n \to\infty} \frac{\log_2 S(n,\floor{\delta n})}{n},\\
\sigma(\delta,\omega) &:=\limsup_{n \to\infty} \frac{\log_2 S(n,\floor{\delta n},\floor{\omega n})}{n}
\end{align*}

First, we reduce the problem of determining $\mu(\delta,\omega)$ to problem
of determining $\alpha(\delta,\omega)$.

\begin{lem}\label{lem:BE:MCWC}
We have $$A(nm,d,mw)\le \frac{\binom{mn}{mw}}{\binom{n}{w}^m} M(m,n,d,w).$$
\end{lem}

Lemma \ref{lem:BE:MCWC} is analogous to Elias-Bassalygo\cite[Theorem 33, Chapter 17]{MacWilliamsSloane:1977} 
by regarding the set of $m$ by $n$ matrices with constant row weight $w$
as a subset of the set of words of length $mn$ with constant-weight $mw$.
As the proof requires some graph theoretical techniques, 
its proof is deferred to Section \ref{sec:BE:MCWC}.

\begin{prop} \label{prop:asymplower}
We have $$\alpha(\delta,\omega) \le \mu(\delta,\omega).$$
\end{prop}
\begin{IEEEproof}
Observe that
\begin{equation*} 
\lim_{n\to\infty} \log \frac{\binom{mn}{mw}}{\binom{n}{w}^m}= mn H(\omega)-mn H(\omega)=0.
\end{equation*}
Then applying limits on $n,m$ and taking logarithms for Lemma \ref{lem:BE:MCWC},
we have our result.
\end{IEEEproof}

The asymptotic version of Proposition \ref{triv} is then

\begin{prop} \label{trivasymp}We have $$\mu(\delta,\omega)\le \alpha(\delta,\omega).$$
\end{prop}

The proof is immediate and omitted. Combining both Propositions \ref{prop:asymplower} and \ref{trivasymp},
we have that the asymptotic exponent of $M(m,n,d,w)$
is equal to the asymptotic exponent of $A(mn,d,mw)$.

\begin{prop}
We have $$\mu(\delta,\omega)= \alpha(\delta,\omega).$$
\end{prop}

Unfortunately, the value of $\alpha(\delta,\omega)$ is in general not known.
Estimates of $\alpha(\delta,\omega)$ are provided by 
McEliece et al. \cite{MRRW:1977} and
Ericson and Zinoviev \cite{EricsonZinoviev:1987}.
In the following subsection, we focus on the case where $\omega=\frac 12$ and
evaluate the asymptotic behavior of the constructions given in 
Section \ref{ssec_cc}.

\subsection{Asymptotics for $\omega=\frac 12$}

The next result follows from the best known upper bound on $\alpha(\delta,\omega)$ due to McEliece {\em et al}.

\begin{prop}[McEliece {\em et al.} {\cite[eq. (2.16)]{MRRW:1977}}]\label{upperbest} 
We have $\mu(\delta,\omega)\le g(u^2),$ with $g(x)=H((1-\sqrt{1-x})/2),$ and
$$ u=-\delta+\sqrt{\delta^2-2\delta+4\omega(1-\omega)}.$$

In particular,
\begin{equation}
\mu(\delta,1/2)\le H(1/2-\sqrt{\delta(1-\delta)}.
\end{equation}
\end{prop}


Our first construction is based on Proposition \ref{lb1}, 
using geometric Goppa codes as outer codes.
In particular, fix $q$ to be a prime power and a square, and fix $0\le \delta \le 1-\frac{1}{\sqrt{q}-1}$.
Tsfasman {\em et al.} \cite{TVZ:1982} 
exhibited the existence of a family of geometric codes with relative distance $\delta$ and rate
\begin{equation*}
\alpha_q(\delta)\ge 1-\delta-\frac{1}{\sqrt{q}-1}.
\end{equation*}


Suppose we pick a CWC$(n,d,n/2)$ of size $q$ as the inner code.
For the outer code, we pick a Goppa $(m,\floor{\delta mn/d})_q$ code of rate at least 
$1-n\delta/d-1/(\sqrt q-1)$. 
Applying Proposition \ref{lb1}, we obtain an MCWC$(m,n, \floor{\delta mn}, n/2)$ of size at least
\begin{equation*}
q^{m(1-n\delta/d-1/(\sqrt q-1))}
\end{equation*}
 
 Taking logarithm, we have our first lower bound for $\mu(\delta,1/2)$.
\begin{thm}\label{lbasym1}
If there exists a CWC$(n,d,n/2)$ of size $q$, then
for $\delta\le d/n(1-1/(\sqrt q-1))$,
$$\mu(\delta,1/2)\ge \frac{\log q}{d}\left(\frac{d}{n}\left(1-\frac {1}{\sqrt q -1}\right)-\delta\right) .$$
\end{thm}

Searching through the online table of lower bounds for $A(n,d,w)$ \cite{Brouwer},
we pick the following constant-weight codes as inner codes:
\begin{enumerate}
\item a CWC$(12,4,6)$ of size $11^2$,
\item a CWC$(28,14,14)$ of size $7^2$,
\item a CWC$(28,4,14)$ of size $1237^2$.
\end{enumerate}

Applying Theorem \ref{lbasym1}, we have
\begin{align}
\mu(\delta,1/2) & \ge \frac{\log 11}{6}\left(\frac{3}{10}-\delta\right),\\
\mu(\delta,1/2) & \ge \frac{\log 7}{14}\left(\frac{5}{12}-\delta\right),\\
\mu(\delta,1/2) & \ge \frac{\log 1237}{14}\left(\frac{1235}{8652}-\delta\right).
\end{align}

\vskip 5pt

Our next construction makes use of the pseudo-product code construction given by Proposition \ref{sys}.
The asymptotic version of this proposition is as follows.

\begin{prop}\label{sysasymp} We have $$\mu(\delta,\omega)\ge \sigma(\delta_1,\omega) \sigma(\delta_2),$$ 
where $0<\delta_1,\delta_2<1$ with $\delta=\delta_1\delta_2$.
\end{prop}

\begin{thm}  \label{lbasym2}
We have for $\delta \le 1/4,$ 
\begin{equation}
\mu(\delta,1/2)\ge (1-H(\sqrt{\delta}))^2/2.
\end{equation}
\end{thm}

\begin{IEEEproof}
By applying Varshamov-Gilbert (VG) bound \cite[Theorem 30, Chapter 17]{MacWilliamsSloane:1977}  to 
systematic codes, we get
$$\sigma(\delta_2)\ge 1-H(\delta_2).$$

Combining VG bound for linear codes with Proposition IV.3 we get

$$\sigma(\delta_1,1/2)\ge (1-H(\delta_1))/2.$$

Using Proposition \ref{sysasymp} with $\delta_1=\delta_2=\sqrt{\delta},$ the result follows.
\end{IEEEproof}

Our final construction follows from setting $q=2$ in Proposition \ref{qary}.

\begin{thm}\label{lbasym3}
We have for $\delta\le 1/2$,
\begin{equation}\label{eq:lb3}
\mu(\delta,1/2)\ge 1-H(\delta).
\end{equation}
\end{thm}

\begin{IEEEproof}
Setting $q=2$ in Proposition \ref{qary} and applying VG bound,
 we have 
 \begin{equation*}
 M(m,2w,2d,w)\ge A(mw,d)\ge 2^{mw(1-H(d/mw))}.
 \end{equation*}
 Taking logarithms, we obtain \eqref{eq:lb3}.
\end{IEEEproof}

Coincidentally, \eqref{eq:lb3} can be obtained directly by observing that
$\mu(\delta,1/2)=\alpha(\delta,1/2)=\alpha(\delta)$.

\begin{figure*}
    \begin{center}
       \includegraphics[width=0.8\linewidth]{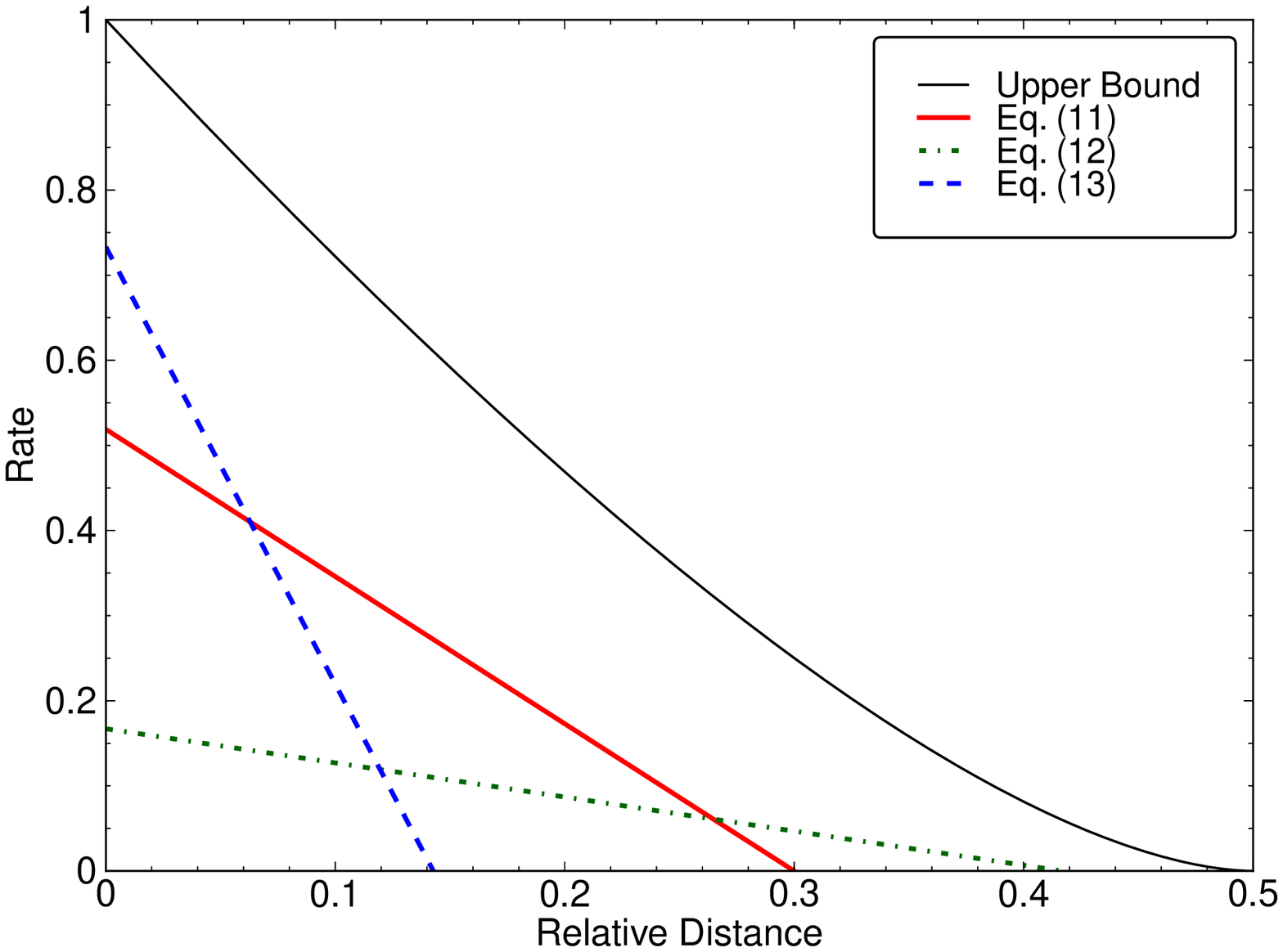}
      
    \end{center}
    
    \begin{center}
       \includegraphics[width=0.8\linewidth]{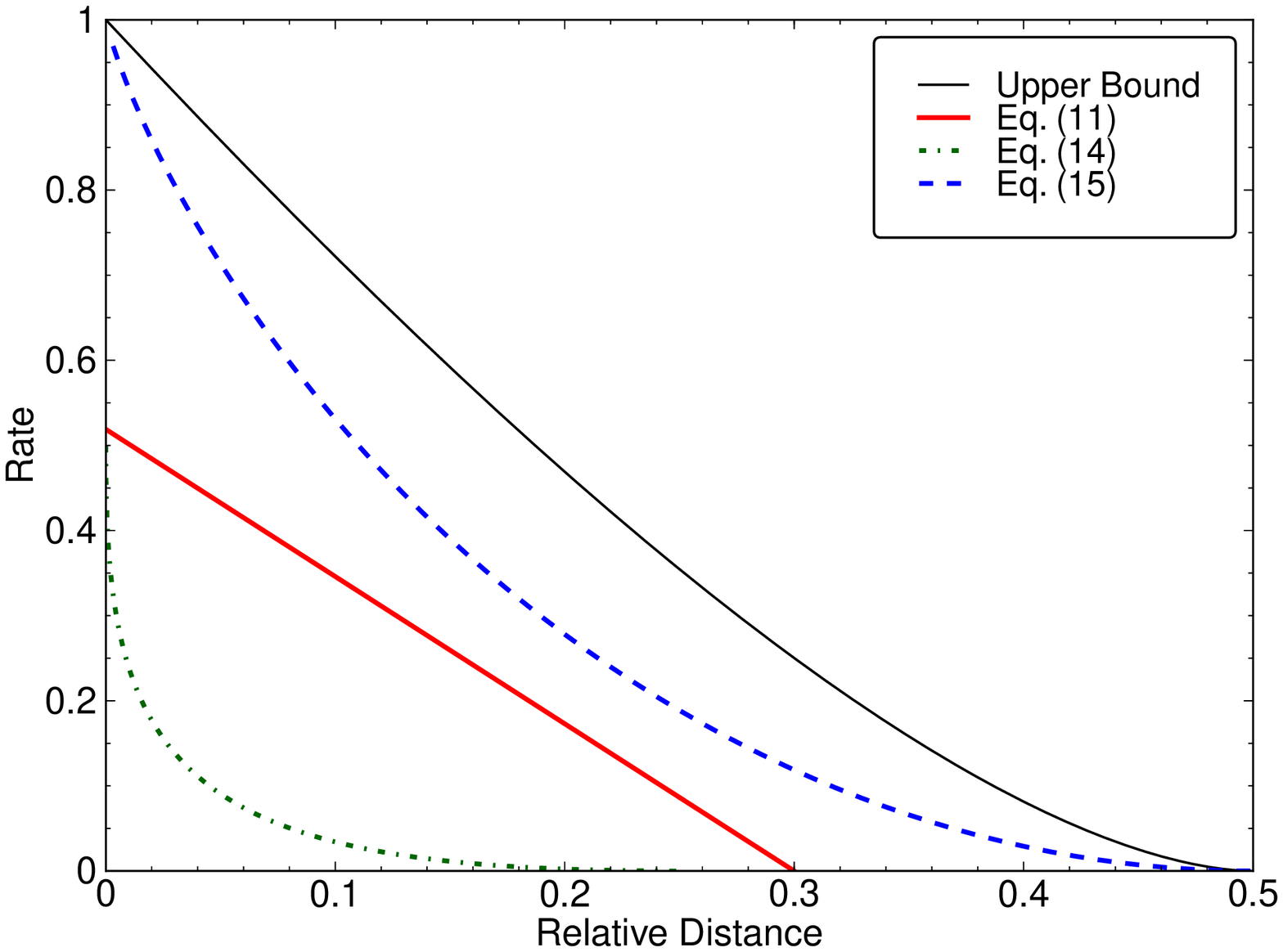}
    \end{center}

\caption{Upper and lower bounds for $\omega=1/2$.}
    \label{fig_bounds}
\end{figure*}

\vskip 5pt

We summarize all the constructions given in this subsection in Figure~\ref{fig_bounds}.
The top graph compares the lower bounds resulting from Theorem \ref{lbasym1} with
various constant-weight codes as inner codes, while 
the bottom graph compares the lower bounds resulting from Theorem \ref{lbasym1}, Theorem \ref{lbasym2} and Theorem \ref{lbasym3}.
We observe that the construction given by Proposition \ref{qary} (or Theorem \ref{lbasym3}) provides the best lower bound.



\subsection{Proof of Lemma \ref{lem:BE:MCWC}}\label{sec:BE:MCWC}

El Rouayheb and Georghiades \cite{ElRouayhebGeorghiades:2012} 
generalized the methods of Elias-Bassalygo 
using graph theoretical methods. 
Below we introduce certain concepts necessary for the proof of Lemma \ref{lem:BE:MCWC}.

Given two graphs $G=(V_G, E_G)$ and $H=(V_H, E_H)$,  
a mapping $\phi : V_G \to V_H$ is called a {\em graph homomorphism} 
if $u,v$ are adjacent in $G$ implies that $\phi(u),\phi(v)$ are adjacent in $H$.
When $G=H$ and $\phi$ is a bijection, then $\phi$ is called an {\em automorphism} of $G$.
Observe that the set of all automorphisms of $G$ is a group under composition; it is called
the {\em automorphism group} of $G$.
A graph is then {\em vertex transitive} 
if the action of its automorphism group on its vertex set is transitive.

Given a graph $G$, a subset $X$ of the vertices is said to be {\em independent}
if every pair of vertices in $X$ is not adjacent in $G$. 
The {\em independence number} of $G$, denoted by $\alpha(G)$, 
the maximum size of an independent set in $G$. 
The following theorem gives the relation between the independence numbers 
of two graphs that are related by a graph homomorphism (see also \cite[Chapter 7]{GodsilRoyle:2001}).

\begin{thm}[El Rouayheb and Georghiades{\cite[Theorem 4]{ElRouayhebGeorghiades:2012}}] \label{thm:graph}
If $H$ is vertex transitive and there is a graph homomorphism from $G$ to $H$,
 then
 \begin{equation*}
\alpha(H)\le \frac{V(H)}{V(G)}\alpha(G).
 \end{equation*}
\end{thm}

Therefore, Lemma \ref{lem:BE:MCWC} is a straightforward application of Theorem \ref{thm:graph}.
Let $G$ be the graph whose vertices are the $m$ by $n$ arrays with constant row weight $w$
and two vertices are adjacent if the distance between the corresponding arrays are less than $d$.
It is then not difficult to observe that an independent set in $G$ corresponds to a multiply constant-weight code of distance $d$
and hence, $\alpha(G)=M(m,n,d,w)$.

Similarly, let $H$ be the graph whose vertices are codewords of length $mn$ with constant row weight $mw$
and two vertices are adjacent if the distance between the corresponding arrays are less than $d$.
We also have $\alpha(H)=A(mn,d,mw)$.

Finally, observe that $G$ is a subgraph of $H$ and hence, we have a graph homomorphism from $G$ to $H$.
Since $H$ is vertex transitive, we apply Theorem \ref{thm:graph} to obtain Lemma \ref{lem:BE:MCWC}.

\section{Conclusion}
\label{sec-conclusion}

Motivated by PUFs, we introduced a new class of codes, called multiply constant-weight codes,
that generalizes constant-weight codes and doubly constant-weight codes.
Using known $q$-ary codes and constant-weight codes as ingredients, 
we construct families of multiply constant-weight codes.
We also provide analogues of the Johnson bound and 
show that the bound is asymptotically tight up to a constant factor, assuming certain conditions.
We then demonstrate that the asymptotic rates of multiply constant-weight codes
and constant-weight codes are the same. 
An analysis of the asymptotic rates of our code constructions are also given.

Finally, we remark that the tabulating the estimates of $M(m,n,d,w)$ for modest values of the four parameters is a worthwhile project.
In addition, the function $S(n,d,w)$ is also worth tabulating and has other applications \cite{BoinckvanTilborg:1990,Lin:1993}.


\section*{Acknowledgements} 
The authors thank Dr. Son Hoang Dau for pointing out the relevant literature 
and Dr. Punarbasu Purkayastha for the helpful discussions.
\bibliographystyle{IEEEtran}
\bibliography{mybibliography}

\end{document}